\documentclass[11pt]{article}

\usepackage[margin=1in]{geometry}
\usepackage{amsmath,amssymb,amsthm}
\usepackage{booktabs}
\usepackage{subcaption}
\usepackage{caption}
\usepackage{graphicx}
\usepackage{setspace}
\usepackage{natbib}
\usepackage[plain,noend]{algorithm2e}

\newtheorem{theorem}{Theorem}
\newtheorem{condition}{Condition}
\newtheorem{remark}{Remark}

\begin{document}

% \title{Theoretical Analysis of Engression and Reverse Markov Engression}

% \author{
% Jiaqi Huang \and Gongjun Xu \and Ji Zhu\\
% Department of Statistics, University of Michigan\\
% \texttt{jiaqihua@umich.edu}, \texttt{gongjun@umich.edu}, \texttt{jizhu@umich.edu}
% }

% \date{}
% \maketitle

% \jname{Biometrika}
% %% The year, volume, and number are determined on publication
% \jyear{2025}
% \jvol{112}
% \jnum{1}
% \cyear{2025}
%% The \doi{...} and \accessdate commands are used by the production team
%\doi{10.1093/biomet/asm023}
% \accessdate{Advance Access publication on 14 February 2025}

%% These dates are usually set by the production team
% \received{2 January 2024}
% \revised{3 February 2025}

%% The left and right page headers are defined here:
% \markboth{J. Huang et~al.}{Biometrika style}

% %% Here are the title, author names and addresses
% \title{Theoretical Analysis of Engression and  Reverse Markov Engression}

% \author{JIAQI HUANG}
% \affil{Department of Statistics, University of Michigan,\\ Ann Arbor, MI 48109, U.S.A.
% \email{jiaqihua@umich.edu}}

% \author{GONGJUN XU}
% \affil{Department of Statistics, University of Michigan,\\ Ann Arbor, MI 48109, U.S.A.
% \email{gongjun@umich.edu}}

% \author{JI ZHU}
% \affil{Department of Statistics, University of Michigan,\\ Ann Arbor, MI 48109, U.S.A.
% \email{jizhu@umich.edu}}
\title{\vspace{-1.5cm}Theoretical Analysis of Engression and Reverse Markov Engression}

\author{
Jiaqi Huang, Gongjun Xu, and Ji Zhu\\[0.3em]
Department of Statistics, University of Michigan\\
Ann Arbor, MI 48109, U.S.A.\\
\texttt{jiaqihua@umich.edu}, \texttt{gongjun@umich.edu}, \texttt{jizhu@umich.edu}
}

\date{}
\maketitle

% \begin{abstract}
% In this paper, we establish theoretical guarantees for Engression and its Reverse Markov extension in conditional distribution learning. Under deep ReLU network parameterizations, we first derive nonasymptotic error bounds for the discrepancy between the distribution generated by the learned Engression generator and the target distribution. To analyze the multi-step Reverse Markov framework, we further develop an Energy-Distance-based chain rule that enables a rigorous error propagation analysis across sequential reverse steps. These results yield near-optimal convergence rates, up to a logarithmic factor, for both Engression and its Reverse Markov extension, thereby matching the minimax optimal rate for nonparametric regression over a general H\"{o}lder class up to logarithmic terms.
% \end{abstract}
\begin{abstract}
Engression is a recently proposed and effective framework for conditional distribution learning. Its multi-step Reverse Markov extension further improves generative flexibility by decomposing complex conditional sampling into sequential reverse transitions. Despite their strong empirical performance, rigorous finite-sample statistical guarantees for these methods remain unavailable. In this paper, under deep neural network parameterizations, we establish nonasymptotic convergence bounds for Engression by directly controlling the Energy Distance between the learned and target conditional distributions. For the Reverse Markov framework, we further develop an Energy-Distance-based chain rule that enables a rigorous analysis of error propagation across reverse steps. Our analysis yields corresponding excess-risk bounds that are near-optimal up to logarithmic factors relative to the classical minimax rate over a general H\"older class.
\end{abstract}

\section{Introduction}
Generative models have been extensively studied for learning and sampling from complex data distributions. Their applications span diverse fields, including image synthesis \citep{pinaya2022brain, song2026wasserstein, gao2020learning}, climate modeling \citep{shen2025reverse}, and the generation of synthetic electronic health records \citep{tian2024reliable} driven by growing privacy concerns. A wide variety of generative model architectures have been proposed, such as diffusion models \citep{sohl2015deep, song2019generative, ho2020denoising, han2022card} and generative adversarial networks (GANs) \citep{zhou2023deep, song2026wasserstein, su2026wasserstein}.

Recently, \cite{shen2025engression} proposed a computationally efficient generative approach named ``Engression''. Engression combines neural networks with the energy score and uses a neural generator to directly learn the conditional distribution of the response from covariates and random noise.
 Building on this, \cite{shen2025reverse} extended the framework with a reverse Markov process to learn complex distributions and provided an efficient discretization strategy for both training and inference. The Engression framework has attracted significant attention and found applications in various fields, with adaptations and extensions to suit specific task requirements, such as image generation and robotics tasks \citep{de2025distributional}, causal margin modelling \citep{yang2025frugal}, and generative time-series modeling for environmental applications \citep{kraft2026modeling}. 

\begin{figure}[htbp]
    \centering
    \includegraphics[width=0.7\linewidth]{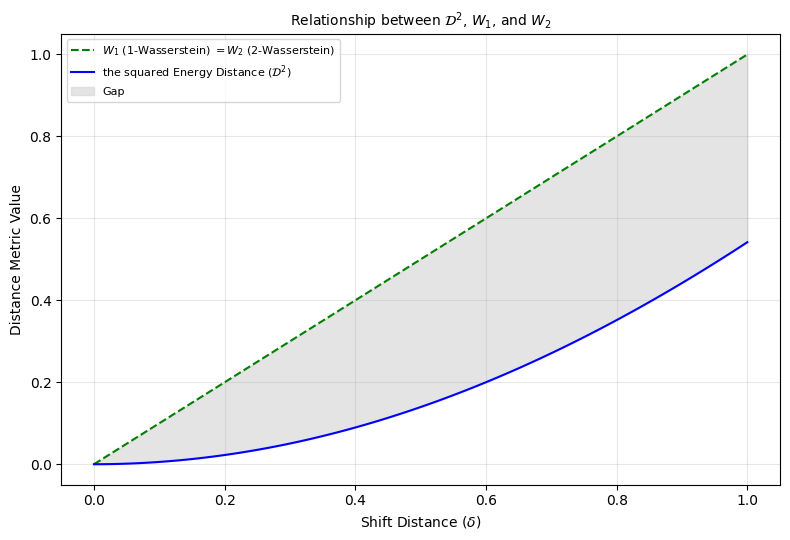}
    \caption{Comparison of distributional metrics under a mean-shift perturbation $\delta$ between $\mathcal{N}(0, 1)$ and $\mathcal{N}(\delta, 1)$. The blue solid line with circular markers represents the squared Energy Distance ($\mathcal{D}^2$). The green dashed line represents the 1-Wasserstein ($W_1$)  and 2-Wasserstein ($W_2$) distances. The light gray shaded area denotes the numerical gap between $\mathcal{D}^2$ and $W_1$.}
    \label{Wassertein}
\end{figure}

Existing studies on Engression and its related Reverse Markov extensions have demonstrated strong empirical performance and established the population validity of the underlying energy-score objectives. However, a rigorous finite-sample error analysis quantifying the discrepancy between the learned and target distributions under neural network approximation is still lacking. 
\citet{shen2025engression} showed
that distributional fitting can improve nonlinear extrapolation, particularly
under preadditive-noise models, while their analysis is mainly focused on
extrapolation and some specific finite-sample settings. 
\cite{shen2025reverse} considered an analysis  through a multi-step framework, but their derivation relies on a restrictive assumption that the Wasserstein approximation error can be explicitly controlled at each discretization step. 
Note that Engression is based on the Energy Distance rather than the Wasserstein distance. 
% Importantly, controlling the Energy Distance does not, in general, imply control of the Wasserstein distance. 
Although it is known that the squared energy-based discrepancy can be bounded from above by the Wasserstein distance \citep{bottou2018geometrical}, this one-way relationship does not yield the converse implication. 
Therefore, theoretical guarantees stated in terms of stepwise Wasserstein errors are not naturally justified by the energy-score objective optimized in Engression. 
To further illustrate this mismatch, Figure~\ref{Wassertein} compares the Energy Distance and Wasserstein distances under a simple mean-shift perturbation. 
The two metrics exhibit different sensitivities to local distributional changes, suggesting that a well-controlled energy-based objective may still correspond to a non-negligible Wasserstein error. 
This motivates developing a finite-sample theory directly under the Energy Distance, the same discrepancy that underlies the Engression objective.

% ??? Challenges.( Rewrite )
% Developing such a theoretical analysis is technically challenging for two main reasons. First, to the best of our knowledge, this is the first work to establish finite-sample convergence rates for neural-network-based generators under an energy-based discrepancy, which requires a nontrivial combination of energy-score analysis and neural network approximation theory. Second, unlike the Wasserstein distance, the energy-based discrepancy does not admit a readily available compositional ``chain rule'' for sequential generative mappings. Consequently, new analytical tools are needed to characterize how statistical errors propagate through successive reverse Markov steps.
Developing a finite-sample theory under the Energy Distance presents two main
challenges. First, existing theories do not directly cover our setting. The
original Engression analysis mainly focuses on extrapolation and specific model
classes, while existing theories for other deep generative models, such as GANs
\citep{chen2020distribution,song2026wasserstein}
and flow matching \citep{zhou2025error}, are typically tailored to different
training objectives, sampling mechanisms, and probability metrics. These existing approaches
cannot be directly applied to the conditional generator trained by the energy-score loss. New analysis is therefore needed to handle the pairwise-distance structure of the Energy-Distance objective and the auxiliary-noise randomness. Second, the reverse Markov framework creates a sequential error-propagation problem, in which stepwise estimation errors may accumulate along the reverse chain. Unlike Wasserstein-type analyses, where pushforward
stability often provides a natural tool for tracking such errors, the Energy
Distance has no direct analogous propagation mechanism. Therefore, quantifying
the accumulation of local Energy-Distance errors requires a dedicated
Energy-Distance-based propagation analysis.

The main contributions of this paper are threefold. 

% First, we establish a finite-sample convergence theory for Engression by quantifying the approximation and estimation errors induced by deep neural network generators. We derive an explicit nonasymptotic convergence bound for the discrepancy between the learned and target distributions, while the induced Energy Distance ($\mathcal{D}^2$) attains the classical near-optimal nonparametric benchmark up to logarithmic factors.

First, we establish a finite-sample convergence theory for Engression by quantifying both the approximation and estimation errors induced by deep neural network generators. We derive an explicit nonasymptotic convergence bound for the Energy Distance between the learned and target conditional distributions. The proof also yields an excess-risk bound that is near-optimal up to logarithmic factors relative to the classical minimax rate over a general H\"older class.

Second, we identify and address a key technical obstacle in the analysis of multi-step generative procedures: the absence of a usable compositional error propagation mechanism under the energy-based discrepancy. To overcome this difficulty, we develop an Energy-Distance-based error propagation inequality, which provides a new analytical tool for controlling the accumulation of statistical errors across sequential reverse transitions.

Finally, building on this chain rule, we extend the finite-sample analysis to the Reverse Markov Engression framework and derive an explicit global convergence bound for the resulting multi-step generator. The bound shows how stepwise errors accumulate along the reverse chain and how the overall rate is governed by the least smooth reverse transition.

\section{Engression and Reverse Markov Learning Frameworks}
Consider the regression setting where we aim to learn the conditional distribution of a response variable $\boldsymbol{Y}\in \mathbb{R}^q$ given covariates $\boldsymbol{X}\in \mathbb{R}^p$.
In this framework, we consider a general stochastic generator class
$\mathcal{M}=\{g(\boldsymbol{x},\boldsymbol{\varepsilon})\},$
where each generator
$g: \mathbb{R}^p \times \mathbb{R}^d \to \mathbb{R}^q$,
$\boldsymbol{x} \in \mathbb{R}^p$ denotes the input covariates, and
$\boldsymbol{\varepsilon} \in \mathbb{R}^d$ is a random noise vector drawn
from a prescribed distribution independent of $\boldsymbol{X}$.
For a fixed $\boldsymbol{x}$, the randomness in
$\boldsymbol{\varepsilon}$ allows $g(\boldsymbol{x}, \boldsymbol{\varepsilon})$
to generate samples from a candidate conditional distribution for
$\boldsymbol{Y}\mid \boldsymbol{X}=\boldsymbol{x}$.

To evaluate such distributional predictions, Engression adopts the energy score \citep{gneiting2007strictly}, which for a predictive distribution $P$ on $\mathbb{R}^q$ and an observation $\boldsymbol{y}$ is defined as
$\mathrm{ES}(P,\boldsymbol{y})= \mathbb{E}_{\boldsymbol{Z},\boldsymbol{Z}'\sim P}\|\boldsymbol{Z}-\boldsymbol{Z}'\|_2/2-
\mathbb{E}_{\boldsymbol{Z}\sim P}\|\boldsymbol{Z}-\boldsymbol{y}\|_2,
$
where $\boldsymbol{Z}$ and $\boldsymbol{Z}'$ are two independent draws from $P$ and $\|\cdot\|_2$ denotes the Euclidean norm. Based on the energy score, \cite{shen2025engression} defines the population Engression solution as
$g^{*}\in\arg\min_{g\in\mathcal{M}}\mathcal{L}(g),
$
where
\begin{equation}\label{engression}
\mathcal{L}(g) = \mathbb{E}\Big\{ \|\boldsymbol{Y} - g(\boldsymbol{X},\boldsymbol{\varepsilon})\|_2 - \frac{1}{2} \|g(\boldsymbol{X},\boldsymbol{\varepsilon}) - g(\boldsymbol{X},\boldsymbol{\varepsilon}')\|_2 \Big\},
\end{equation}
and $\boldsymbol{\varepsilon},\boldsymbol{\varepsilon}'$ are independent draws from the noise distribution. Proposition 1 of \cite{shen2025engression} shows that under a realizability condition on the generator class $\mathcal{M}$, the population Engression solution $g^*$ recovers the target conditional distribution almost everywhere.
In practice, we approximate this target by optimizing over a class of neural network-based functions, denoted by $\mathcal{G}$, such as deep neural networks with ReLU activation functions.

Let $\{(\boldsymbol{X}_i, \boldsymbol{Y}_i)\}_{i=1}^n$ 
be i.i.d. samples drawn from the joint distribution of $(\boldsymbol{X},\boldsymbol{Y}).$ 
For each observation $\boldsymbol{X}_i$, we generate $m$ independent noise samples 
$\{\boldsymbol{\varepsilon}_{ij}\}_{j=1}^m$ from the noise distribution. 
The resulting empirical estimator is then defined as $\hat{g} \in \arg\min_{g \in \mathcal{G}} \hat{\mathcal{L}}(g),$
where the empirical risk $\hat{\mathcal{L}}(g)$ is given by
\begin{equation*}\label{empiricalloss}
    \hat{\mathcal{L}}(g) = \frac{1}{n} \sum_{i=1}^{n} \Bigg\{ \frac{1}{m} \sum_{j=1}^{m} \|\boldsymbol{Y}_i - g(\boldsymbol{X}_i, \boldsymbol{\varepsilon}_{ij})\|_2 - \frac{1}{2m(m-1)} \sum_{j=1}^{m} \sum_{\substack{j' = 1 \\j' \neq j}}^{m} \|g(\boldsymbol{X}_i, \boldsymbol{\varepsilon}_{ij}) - g(\boldsymbol{X}_i, \boldsymbol{\varepsilon}_{ij'})\|_2 \Bigg\}.
\end{equation*}

To learn highly complex conditional distributions, a single-step 
generator may be insufficient to capture intricate structures. 
In particular, directly mapping noise to data in one step can be 
challenging, often leading to overly smooth approximations and 
unrealistic samples in low-density regions \citep{shen2025reverse}. To address this limitation, 
\cite{shen2025reverse} proposed combining Engression with a multi-step 
Reverse Markov Process, which decomposes the generation task into a 
sequence of simpler conditional transformations.
Their approach bridges the unknown target conditional distribution \(P_{g^*}(\cdot \mid \boldsymbol{X}=\boldsymbol{x})\) to a known reference conditional distribution \(Q^*(\cdot \mid \boldsymbol{X}=\boldsymbol{x})\) over a finite number of steps $S$, such as diffusion, linear interpolation, or spatial pooling, producing a conditional stochastic process $\{\boldsymbol{Y}^0 = \boldsymbol{Y}, \boldsymbol{Y}^1, \dots, \boldsymbol{Y}^S\}\mid \boldsymbol{X}=\boldsymbol{x}$. This forward process gradually transforms the target distribution 
into a simpler reference distribution through a sequence of 
intermediate states. For instance, in a diffusion-based forward process following \cite{ho2020denoising}, we have
$
\boldsymbol{Y}^s \mid \boldsymbol{Y}^{s-1} \sim \mathcal{N}(\sqrt{1 - \sigma_s} \boldsymbol{Y}^{s-1}, \sigma_s^2 I),$
with an increasing schedule $\sigma_s$.

In the reverse process, each conditional Markov distribution is learned via Engression. Specifically, for each step $s$, let \(P_{g_{(s)}^*}(\cdot \mid \boldsymbol x,\boldsymbol y^s)\) denote the target conditional distribution of \(\tilde{\boldsymbol Y}^{s-1}\) given \((\boldsymbol X,\tilde{\boldsymbol Y}^s)=(\boldsymbol x,\boldsymbol y^s)\), where $\{\tilde{\boldsymbol{Y}}^s\}_{s=0}^S$ is generated by the reverse 
Markov chain, which sequentially reconstructs samples from the 
target conditional distribution by iteratively transforming samples 
from the reference distribution backward through the learned 
conditional mappings. For each step $s\in \{1,2,\ldots,S\}$, consider a general stochastic generator class
$\mathcal{M}^s=\{g(\boldsymbol{x},\boldsymbol{y}^s,\boldsymbol{\varepsilon}_s)\},$
where $g:(\boldsymbol{x},\boldsymbol{y}^s,\boldsymbol{\varepsilon}_s)\mapsto \boldsymbol{y}^{s-1}$ belongs to a function class and $\boldsymbol{\varepsilon}_s$ is a random noise vector with a prescribed distribution independent of $(\boldsymbol{X},\boldsymbol{Y}^s)$. 
We define the stepwise population Engression solution as
$
g_{(s)}^* \in \arg\min_{g\in \mathcal{M}^s} \mathcal{L}_s(g),
$
where
$$\mathcal{L}_s(g)= \mathbb{E}\Big\{ \|\boldsymbol{Y}^{s-1} - g(\boldsymbol{X}, \boldsymbol{Y}^s, \boldsymbol{\varepsilon}_s)\|_2 - \frac{1}{2} \|g(\boldsymbol{X}, \boldsymbol{Y}^s, \boldsymbol{\varepsilon}_s) - g(\boldsymbol{X}, \boldsymbol{Y}^s, \boldsymbol{\varepsilon}_s')\|_2 \Big\}.$$
Its empirical counterpart is defined as
$\hat{g}_{(s)} \in \arg\min_{g\in \mathcal{G}_s} \hat{\mathcal{L}}_s(g)$,
where $\mathcal{G}_s$ denotes the class of deep ReLU neural networks at step $s$, and (with denoting $\boldsymbol{Z}_i^s = (\boldsymbol{X}_i, \boldsymbol{Y}_i^s)$)
\begin{align*}
    \hat{\mathcal{L}}_s(g) = \frac{1}{n} \sum_{i=1}^{n} \Bigg\{ \frac{1}{m} \sum_{j=1}^{m} \|\boldsymbol{Y}_i^{s-1} - g(\boldsymbol{Z}_i^s, \boldsymbol{\varepsilon}_{ij}^s)\|_2 - \frac{1}{2m(m-1)} \sum_{j=1}^{m} \sum_{\substack{j' = 1 \\ j' \neq j}}^{m} \|g(\boldsymbol{Z}_i^s, \boldsymbol{\varepsilon}_{ij}^s) - g(\boldsymbol{Z}_i^s, \boldsymbol{\varepsilon}_{ij'}^s)\|_2 \Bigg\}.
\end{align*}

% For each step $s\in \{1,2,\ldots,S\}$, we define
% $g_{(s)}^* \in \arg\min_{g\in \mathcal{M}^s} \mathcal{L}_s(g),$ where $\mathcal{M}^s=\{g(\boldsymbol{x},\boldsymbol{y}^s,\varepsilon_s)\}
% $
% and 
% $$\mathcal{L}_s(g)= \mathbb{E}\Big\{ \|\boldsymbol{Y}^{s-1} - g(\boldsymbol{X}, \boldsymbol{Y}^s, \boldsymbol{\varepsilon}_s)\|_2 - \frac{1}{2} \|g(\boldsymbol{X}, \boldsymbol{Y}^s, \boldsymbol{\varepsilon}_s) - g(\boldsymbol{X}, \boldsymbol{Y}^s, \boldsymbol{\varepsilon}_s')\|_2 \Big\},$$
% with the empirical counterpart $
% \hat{g}_{(s)} \in \arg\min_{g\in \mathcal{G}_s} \hat{\mathcal{L}}_s(g),$
% where $\mathcal{G}_s$ denotes the  class of deep ReLU neural networks in $s$th step and denoting  $\boldsymbol{Z}_i^s = (\boldsymbol{X}_i, \boldsymbol{Y}_i^s)$,
% \begin{align*}
%     \hat{\mathcal{L}}_s(g) = \frac{1}{n} \sum_{i=1}^{n} \Bigg\{ \frac{1}{m} \sum_{j=1}^{m} \|\boldsymbol{Y}_i^{s-1} - g(\boldsymbol{Z}_i^s, \boldsymbol{\varepsilon}_{ij}^s)\|_2 - \frac{1}{2m(m-1)} \sum_{j=1}^{m} \sum_{\substack{j' = 1 \\ j' \neq j}}^{m} \|g(\boldsymbol{Z}_i^s, \boldsymbol{\varepsilon}_{ij}^s) - g(\boldsymbol{Z}_i^s, \boldsymbol{\varepsilon}_{ij'}^s)\|_2 \Bigg\}.
% \end{align*}
Finally, given a new observation $\boldsymbol{x}$, the reverse sampling process proceeds as
$
\tilde{\boldsymbol{y}}^{(s-1)} = \hat{g}_{(s)}(\boldsymbol{x}, \tilde{\boldsymbol{y}}^{(s)}, \boldsymbol{\varepsilon}^s),$ $ s = S, S-1, \dots, 1,
$
where $\tilde{\boldsymbol{y}}^{(S)}$ is drawn from the known distribution $Q^*( \cdot \mid \boldsymbol{X} = \boldsymbol{x})$. This yields a sample $\tilde{\boldsymbol{y}}^{(0)}$ from the estimated conditional distribution.
%\subsection{Implementation}

\section{Model Architectures and Theoretical Results}

The goal of this section is to establish finite-sample guarantees under the same discrepancy that is optimized by the Engression objective. 
Since Engression is trained through the energy score, its population excess risk is naturally linked to the Energy Distance between the generated and target conditional distributions. 
More precisely, Lemma~1 in the Supplementary Material shows that the population excess Engression loss is proportional to the conditional squared Energy Distance, averaged over the covariate distribution. 
Therefore, rather than translating the analysis into a Wasserstein-type error, we study the convergence of the learned conditional distribution directly under the Energy Distance. 
This aligns the theoretical error criterion with the training objective and avoids relying on assumptions that are not implied by the energy-score formulation.
For two probability distributions \(P\) and \(Q\) on \(\mathbb{R}^q\) with finite first moments, we define the Energy Distance \citep{bottou2018geometrical} as
$
\mathcal{D}(P, Q)
:=
\left\{
2\mathbb{E}\|\boldsymbol{U}- \boldsymbol{V}\|_2
-
\mathbb{E}\|\boldsymbol{U} - \boldsymbol{U}'\|_2
-
\mathbb{E}\|\boldsymbol{V} - \boldsymbol{V}'\|_2
\right\}^{1/2},
$ where \(\boldsymbol{U}, \boldsymbol{U}' \overset{\mathrm{i.i.d.}}{\sim} P\) and
\(\boldsymbol{V}, \boldsymbol{V}' \overset{\mathrm{i.i.d.}}{\sim} Q\) are independent. 
Moreover, \(\mathcal{D}\) is a metric on probability distributions and
\(\mathcal{D}(P,Q)=0\) if and only if \(P=Q\); see \citet{szekely2013energy}.

\subsection{Single-step Engression}\label{single}

In this paper, we approximate the true generator $g^*$ 
using neural networks with the ReLU activation function \citep{schmidt2020nonparametric}.
The ReLU function is denoted by $\sigma(\boldsymbol{x}) := \max(\boldsymbol{x},0)$ for each component of $\boldsymbol{x}$ if $\boldsymbol{x}$ is a vector. 
Let $\mathcal{G} \equiv \mathrm{NN}(p+d, q, W_G, H_G)$ be a class of ReLU-activated feedforward neural networks $g_\theta: \mathbb{R}^{p+d} \to \mathbb{R}^q$, with parameter $\theta$, width $W_G$, and depth $H_G$. 
Each network in $\mathcal{G}$ can be expressed as a composite function
$
g_\theta(\boldsymbol{x},\boldsymbol{\varepsilon})
=
L_{H_G} \circ \sigma \circ L_{H_G-1} \circ \sigma \circ \cdots \circ \sigma \circ L_1 \circ \sigma \circ L_0(\boldsymbol{x},\boldsymbol{\varepsilon}),
(\boldsymbol{x},\boldsymbol{\varepsilon}) \in \mathbb{R}^{p+d},
$
where $\circ$ denotes function composition, and 
$L_i(\boldsymbol{z}) = \boldsymbol{W}_i \boldsymbol{z} + \boldsymbol{b}_i$ with a weight matrix 
$\boldsymbol{W}_i \in \mathbb{R}^{p_{i+1} \times p_i}$ and bias vector 
$\boldsymbol{b}_i \in \mathbb{R}^{p_{i+1}}$ in the $i$th linear transformation, and $p_i$ is the width of the $i$th layer, $i = 0,1,\ldots,H_G$. 
The width and depth of the network are described by 
$W_G = \max\{p_1, \ldots, p_{H_G}\}$ and $H_G$, respectively.
% We now specify the function classes used in our framework:
% \begin{itemize}
%     \item[] \textbf{Generator network class for engression $\mathcal{G}$:} 
%     Let $\mathcal{G} \equiv \mathrm{NN}(p+d, q, W_G, H_G)$ be a class of ReLU-activated feedforward neural networks $g_\theta: \mathbb{R}^{p+d} \to \mathbb{R}^q$, with parameter $\theta$, width $W_G$, and depth $H_G$.

%     \item[] \textbf{Generator network class for reverse Markov steps $\mathcal{G}_s$:} 
%     For each step $s = 1,\ldots,S$, let $\mathcal{G}_s \equiv \mathrm{NN}(p+d+q, q, W_G, H_G)$ be a class of ReLU-activated feedforward neural networks $g_{s,\theta}: \mathbb{R}^{p+d+q} \to \mathbb{R}^q$, with parameter $\theta$, width $W_G$, and depth $H_G$. Note that the additional $q$ input dimensions correspond to the conditioning on $\boldsymbol{Y}^s$.
% \end{itemize}

We introduce the following notation and assumptions for our theoretical results. For any $\beta > 0$, $B>0$ and a set $\Omega \subseteq \mathbb{R}^{p+d}$, we define the H\"{o}lder class of functions $\mathcal{H}^\beta(\Omega, B)$
as 
%with constant $0 < B < \infty$ is defined as
\[
\mathcal{H}^\beta(\Omega, B) = \biggr\{ f : \Omega \to \mathbb{R} : \max_{\|\boldsymbol{\alpha}\|_1 \leq \lfloor \beta\rfloor} \|\partial^{\alpha} f\|_\infty \leq B, \max_{\|\boldsymbol{\alpha}\|_1 = \lfloor\beta\rfloor} \sup_{\substack{x,y \in \Omega \\ x \neq y}} \frac{|\partial^\alpha f(x) - \partial^\alpha f(y)|}{\|x - y\|^{\beta - \lfloor\beta\rfloor}} \leq B \biggr\},
\]
where $\|\boldsymbol{\alpha}\|_1=\sum_{i=1}^{p+d}\alpha_i$, $\lfloor\beta\rfloor$ denotes the largest integer strictly smaller than $\beta$, $\partial^\alpha = \partial^{\alpha_1} \cdots \partial^{\alpha_{p+d}}$ with $\boldsymbol{\alpha} = (\alpha_1, \ldots, \alpha_{p+d})^\top \in \mathbb{N}_0^{p+d}$ and $\mathbb{N}_0$ is the union of the set of positive integers and ${0}$. 
Let $\mathcal{X} \subseteq \mathbb{R}^p$ denote the domain of the covariates $\boldsymbol{X}$, $\mathcal{Y} \subseteq \mathbb{R}^q$ the domain of the response $\boldsymbol{Y}$, and $\Omega_\varepsilon \subseteq \mathbb{R}^d$ the domain of the noise variable $\boldsymbol{\varepsilon}$.

\begin{condition}\label{boundedness}
 The probability measures of \((\boldsymbol{X}, \boldsymbol{Y})\) and \((\boldsymbol{X}, g(\boldsymbol{X}, \boldsymbol{\varepsilon}))\) are supported on a compact set \(\mathcal{X} \times \mathcal{Y} \subseteq [-B_1, B_1]^{p+q}\) for any \(g \in \mathcal{G}\), with a constant \(0 < B_1 < \infty\). The noise vector $\boldsymbol{\varepsilon}$ is sub-exponential with parameter $L > 0$. Specifically, for each component $\varepsilon_j$, $j=1,\dots,d$, and any $t > 0$, its tail probability satisfies $  \mathbb{P}(|\varepsilon_j| > t) \leq 2 \exp \left( -t/L \right)$.
% Consequently, for the $d$-dimensional vector, $\mathbb{P}(\|\boldsymbol{\varepsilon}\|_\infty > t) \leq 2d \exp(-t/L)$.
\end{condition}
% \begin{condition}\label{noise}
% The probability measure of \(\boldsymbol{\varepsilon}\) is supported on \(\Omega_{\varepsilon} \subseteq [-B_1, B_1]^d\).
% \end{condition} 
\begin{condition}\label{smoothness}
For the Engression setting, the true generator $g^* = (g_1^*, \dots, g_q^*)^\top$ satisfies $g_k^* \in \mathcal{H}^\beta(\mathcal{X} \times \Omega_\varepsilon, B_1)$ for $k = 1, \dots, q$. 

% For the reverse Markov setting, assume that for each step $s = 1, \dots, S$, the step-specific true generator $g_{(s)}^* = (g_{(s),1}^*, \dots, g_{(s),q}^*)^\top$ satisfies $g_{(s),k}^* \in \mathcal{H}^\beta(\mathcal{X} \times \mathcal{Y}^s \times \Omega_\varepsilon, B_1)$ for $k = 1, \dots, q$.
\end{condition}
% \begin{condition}\label{smoothness_unbounded}
% For the engression setting, we assume the true generator $g^* = (g_1^*, \dots, g_q^*)^\top$ is defined on $\mathcal{X} \times \mathbb{R}^d$. For each component $k = 1, \dots, q$, we require $g_k^* \in \mathcal{H}^\beta(\mathcal{X} \times \mathbb{R}^d, B_1)$, meaning that for any compact subset $K \subset \mathcal{X} \times \mathbb{R}^d$, the restriction of $g_k^*$ to $K$ satisfies the Hölder condition with bound $B_1$.
% \end{condition}
\begin{condition}\label{network} 
% The generator class \(\mathcal{G}\) consists of ReLU neural networks with depth \(H_{\mathcal{G}}\) and width $W_{\mathcal{G}}$.
Let $\bar{W}, \bar{H}\in \mathbb{N}$, which may depend on $n$.
    The generator ReLU network class $\mathcal{G} = \mathrm{NN}(p + d, q, W_G, H_G)$ has width 
$W_G = 38q(\lfloor \beta \rfloor + 1)^2 3^{p+d} (m + d)^{\lfloor \beta \rfloor + 1}\bar{W} \log_2 (8\bar{W})$ 
and depth
$H_G = 21(\lfloor \beta \rfloor + 1)^2 \bar{H}[\log_2 (\bar{H})] + 2(p + d).$ 
\end{condition}

\begin{remark}
Conditions~\ref{boundedness}--\ref{network} are in line with the regularity assumptions commonly adopted in recent theoretical studies of deep neural generative models, including GAN-based distribution approximation and estimation \citep{chen2020distribution,su2026wasserstein}, and flow matching \citep{zhou2025error}. Throughout this subsection, the generator class \(\mathcal{M}\) is studied under the above H\"older-smooth regularity framework.
In particular, Condition~\ref{boundedness} relaxes the bounded-support requirement on the noise that is often used in related theoretical analyses. 
We only require a sub-exponential tail condition, allowing \(\boldsymbol{\varepsilon}\) to have unbounded support on \(\mathbb{R}^d\) while keeping the neural approximation and stochastic error terms controllable.
\end{remark}

We next present the main theoretical result for single-step Engression. 
Our goal is to quantify the discrepancy between the conditional distribution generated by the fitted neural network and the target conditional distribution. 
Recall that \(P_{g^*}(\cdot \mid \boldsymbol X)\) is the target conditional distribution, and let
\(P_{\hat g}(\cdot \mid \boldsymbol X)\) be the distribution induced by the generator
\(\hat g(\boldsymbol X,\boldsymbol\varepsilon)\). 
The estimator \(\hat g\) is obtained by minimizing the empirical Engression objective over \(\mathcal G\), based on
$
\mathcal{S}
=
\left\{(\boldsymbol{X}_i,\boldsymbol{Y}_i)\right\}_{i=1}^n
\cup
\left\{\boldsymbol{\varepsilon}_{ij}\right\}_{i=1,\dots,n;\,j=1,\dots,m}.
$
Thus, \(\hat g\) is random through \(\mathcal S\), and the expectation \(\mathbb E_{\mathcal S}\) averages over repeated ${\mathcal S}$.
%the observed samples and auxiliary noise variables. 
The following theorem gives its finite-sample convergence rate under the Energy Distance.
\begin{theorem}\label{rates}
    Suppose Conditions \ref{boundedness}-\ref{network} hold. Then for some positive constant \(C\) independent of \(n\), we have
    $\mathbb{E}_{\mathcal S}\left[
\mathbb{E}_{\boldsymbol X}
\left[
\mathcal{D}\big(
P_{\hat g}(\cdot\mid \boldsymbol X),
P_{g^*}(\cdot\mid \boldsymbol X)
\big)
\right]
\right]
\le
C n^{-\frac{\beta}{2(p+d+2\beta)}}
\log(n)^{\frac{\beta}{p+d}+\frac{\beta\vee 1}{2}}.$
% $$ \leq Cn^{-\frac{\beta}{p+d+2\beta}}\log(n)^{\frac{2\beta}{p+d}+ 1}$$
% and 
% \begin{equation*}\label{ratessig}

% \end{equation}
%where $C$ is a positive constant independent of $n$.
\end{theorem}

\begin{remark}
The proof of Theorem \ref{rates} also yields the excess risk bound  $$\mathbb E_{\mathcal S}\{\mathcal L(\hat g)-\mathcal L(g^*)\}=\mathcal O\!\left(
n^{-\beta/(p+d+2\beta)}
\log(n)^{1\vee\beta+2\beta/(p+d)}
\right),$$ where \(\mathcal L(\hat g)-\mathcal L(g^*)\) denotes the population Engression loss gap. Up to logarithmic factors, this rate is near-optimal relative to the classical minimax rate for nonparametric regression over a \((p+d)\)-dimensional H\"older class \citep{gyorfi2002distribution}. 
Related finite-sample convergence analyses for other deep generative models have recently been developed; see, for example, \citet{chen2020distribution}, \citet{zhou2025error}, \citet{su2026wasserstein} and \cite{song2026wasserstein}. In our setting, our result is derived directly under the Energy Distance induced by the Engression objective. To the best of our knowledge, this is the first finite-sample statistical estimation guarantee established for Engression under deep neural network parameterization.
This theorem also provides the fundamental local convergence bound used in the subsequent reverse Markov analysis.
\end{remark}
% \begin{remark}
%    Theorem \ref{rates} provides an upper bound on the expected excess risk and, via Lemma 1 in Supplementary Material, on the expected Energy Distance between the generated distribution \(P_{\hat{g}}\) and the target distribution \(P_{g^*}\). The obtained rate \(\mathcal{O}\big(n^{-{\beta}/{(p+d+2\beta)}} \log(n)^{({2\beta}/{(p+d)})+ 1}\big)\) is near-optimal: up to a logarithmic factor, it attains the minimax optimal rate for nonparametric regression in a H\"older class \(\mathcal{H}^{\beta}\) over a \((p+d)\)-dimensional space. This result is consistent with the classical nonparametric theory \citep{gyorfi2002distribution}.  A similar near-optimal rate was obtained by \citet{su2026wasserstein} in the Wasserstein Generative Regression framework. Our result achieves comparable convergence while avoiding the adversarial training dynamics inherent to GAN-based approaches. This theorem also provides a basis for deriving convergence rates within the reverse Markov framework. 

% \end{remark}

\subsection{Reverse Markov Learning Conditions}
This subsection extends the analysis to the multi-step reverse Markov framework. The sequential nature of the process introduces additional complexity, and thus extra conditions are imposed to control error propagation and ensure theoretical guarantees. Specifically,
Conditions \ref{cond:boundedness_markov}-\ref{cond:network_markov} extend the regularity conditions from the single-step Engression setting to the multi-step reverse Markov framework. Condition~\ref{new} imposes an Energy-Distance stability requirement on the true
reverse generators. This condition controls the pullback of RKHS functions
through each reverse transition and serves as the structural foundation for the cumulative error analysis.
% For the reverse Markov setting, let $\mathcal{Y}^s \subseteq \mathbb{R}^q$ denote the domain of the intermediate variable $\boldsymbol{Y}^s$ at step $s$. 
% Following the neural network notation introduced in Subsection~\ref{single} for single-step Engression, for each  $s=1,\ldots,S$, we define
% $
% \mathcal{G}_s \equiv \mathrm{NN}(p+d+q, q, W_G, H_G),
% $
% where $\mathcal{G}_s$ is a class of ReLU-activated feedforward neural networks
% $g_{s,\theta}: \mathbb{R}^{p+d+q} \to \mathbb{R}^q$ with parameter $\theta$, width $W_G$, and depth $H_G$. 
% Each $g_{s,\theta}\in\mathcal{G}_s$ takes $(\boldsymbol{X},\boldsymbol{Y}^s,\boldsymbol{\varepsilon}_s)$ as input and outputs an approximation to $\boldsymbol{Y}^{s-1}$. 
% The additional $q$ input dimensions correspond to the conditioning on the intermediate state $\boldsymbol{Y}^s$. 
Following the neural network notation introduced in Subsection~\ref{single},
for each step $s \in \{1, \dots, S\}$, the generator class $\mathcal{G}_s$ consists of ReLU neural networks with depth $H_{\mathcal{G},s}$ and width $W_{\mathcal{G},s}$. 
For the reverse Markov setting, let $\mathcal{Y}^s \subseteq \mathbb{R}^q$ denote the domain of the intermediate variable $\boldsymbol{Y}^s$ at step $s$.
\begin{condition}
\label{cond:boundedness_markov}
There exists a constant \(0<B_1<\infty\) such that, for each step \(s=1,\ldots,S\), 
the probability measures of
\((\boldsymbol X,\boldsymbol Y^s,\boldsymbol Y^{s-1})\) and
\((\boldsymbol X,\boldsymbol Y^s,g_{(s)}(\boldsymbol X,\boldsymbol Y^s,\boldsymbol\varepsilon_s))\)
are supported on
$
\mathcal X\times\mathcal Y^s\times\mathcal Y^{s-1}
\subseteq[-B_1,B_1]^{p+2q},
$
for any \(g_{(s)}\in\mathcal G_s\). 
For each step \(s=1,\ldots,S\), the noise vector
\(\boldsymbol\varepsilon_s=(\varepsilon_{s,1},\ldots,\varepsilon_{s,d})^\top\)
is zero-mean and satisfies the componentwise sub-exponential tail condition
$
\mathbb P(|\varepsilon_{s,j}|>t)\le 2\exp(-t/L), j=1,\ldots,d$, where $t>0$ and $L>0$ is the scale parameter.
% There exists a constant $0 < B_1 < \infty$ such that the support of the  predictor $\boldsymbol{X}$ satisfies $\mathcal{X} \subseteq [-B_1, B_1]^p$. For each step $s \in \{1, \dots, S\}$, the noise $\boldsymbol{\varepsilon}_s$ is a zero-mean random vector satisfying a sub-exponential condition $\mathbb{P}(| \boldsymbol{\varepsilon}_s | > t) \leq 2 \exp(-t/L), \forall t > 0,$
% where $L > 0$ is the scale parameter. 
% % and the noise $\boldsymbol{\varepsilon}_s$ for each step satisfies $\Omega_{\varepsilon} \subseteq [-B_1, B_1]^p$.
% Furthermore, for each $s \in \{1, \dots, S\}$, the range of both the true reverse generator $g_{(s)}^*$ and the estimated generator $\hat{g}_{(s)}$ are contained within the compact set $\mathcal{Y}^s \subseteq [-B_1, B_1]^q$.
\end{condition}

% \begin{condition}\label{cond:noise_markov}
% The probability measures of the independent noise terms $\{\boldsymbol{\varepsilon}_s\}_{s=1}^S$ are supported on $\Omega_{\varepsilon} \subseteq [-B_1, B_1]^p$. For each $s$, $\boldsymbol{\varepsilon}_s$ is assumed to be i.i.d. following a distribution on $\Omega_{\varepsilon}$.
% \end{condition}

\begin{condition}\label{cond:smoothness_markov}
For each step $s \in\{ 1, \ldots, S\}$, the true generator $g_{(s)}^* = (g_{(s),1}^*, \dots, g_{(s),q}^*)^\top$ satisfies $g_{(s),k}^* \in \mathcal{H}^{\beta_s}(\mathcal{X} \times \mathcal{Y}^s \times \Omega_\varepsilon, B_{1,s})$ for $k = 1, \dots, q$. Here, $\beta_s$ and $B_{1,s}$ represent the smoothness level and the uniform bound for step $s$, respectively.
\end{condition}

\begin{condition}\label{cond:network_markov}
For each step $s \in \{1, \dots, S\}$, let $\bar{W}_s, \bar{H}_s \in \mathbb{N}$ be parameters, which may depend on $n$. The ReLU network class $\mathcal{G}_s = \mathrm{NN}(p + q + d, q, W_{G,s}, H_{G,s})$ for the $s$th generator has width 
$W_{G,s} = 38q(\lfloor \beta_s \rfloor + 1)^2 3^{p+q+d} (p+q+d)^{\lfloor \beta_s \rfloor + 1}\bar{W}_s \log_2 (8\bar{W}_s)$ 
and depth $H_{G,s} = 21(\lfloor \beta_s \rfloor + 1)^2 \bar{H}_s[\log_2 (\bar{H}_s)] + 2(p+q+d).$ 
\end{condition}

% \begin{condition}[Lipschitz Continuity]\label{cond:reverse_lipschitz}For each $s \in \{1, \dots, S\}$, the target reverse model $g_{(s)}^*(\boldsymbol{x}, \boldsymbol{y}^s, \boldsymbol{\varepsilon}_s)$ is $L_s$-Lipschitz with respect to $\boldsymbol{y}^s$:$$\|g_{(s)}^*(\boldsymbol{x}, \boldsymbol{y}^a, \boldsymbol{\varepsilon}_s) - g_{(s)}^*(\boldsymbol{x}, \boldsymbol{y}^b, \boldsymbol{\varepsilon}_s)\|_2 \le L_s \|\boldsymbol{y}^a - \boldsymbol{y}^b\|_2, \quad \forall \boldsymbol{y}^a, \boldsymbol{y}^b \in \mathbb{R}^q.$$
% %We define the cumulative Lipschitz constant as $M_s = \prod_{k=1}^{s-1} L_k$, with $M_1 = 1$.
% \end{condition}
\begin{condition}\label{new}
For each \(s=1,\ldots,S\) and each covariate value \(\boldsymbol{x}\), define
$
T_s^x(\boldsymbol{y},\boldsymbol{\varepsilon}_s)=g_{(s)}^*(\boldsymbol{x},\boldsymbol{y},\boldsymbol{\varepsilon}_s),
 \boldsymbol{y}\in\mathcal Y^{(s)}.
$
Let \(k_{s,\mathrm{out}}\) and \(k_{s,\mathrm{in}}\) be the distance-induced kernels
associated with the Energy Distance on \(\mathcal Y^{(s-1)}\) and
\(\mathcal Y^{(s)}\), respectively, and let
\(\mathcal B_{k_{s,\mathrm{out}}}\) and \(\mathcal B_{k_{s,\mathrm{in}}}\)
denote the corresponding RKHS unit balls. For every
\(f\in \mathcal B_{k_{s,\mathrm{out}}}\), define
$
\ell_{s,f}^x(\boldsymbol{y})
=
\mathbb E_{\varepsilon_s}
\left[
f\{T_s^x(\boldsymbol{y},\boldsymbol{\varepsilon}_s)\}
\right].
$
There exist a constant \(M_s<\infty\) and an anchor point
\(\boldsymbol{y}_0\in\mathcal Y^{(s)}\) such that, for every
\(f\in\mathcal B_{k_{s,\mathrm{out}}}\),
$
\|
\ell_{s,f}^x(\cdot)-\ell_{s,f}^x(\boldsymbol{y}_0)
\|_{\mathcal H_{k_{s,\mathrm{in}}}}
\le
M_s,
$
uniformly over \(\boldsymbol{x}\).
\end{condition}
\begin{remark}
The Energy Distance admits an equivalent maximum mean discrepancy (MMD) representation induced by a
distance-based kernel; see, for example,
\citet{sejdinovic2013equivalence} and \citet{szekely2013energy}. In this
representation, distributional discrepancies are characterized by expectation
differences over the unit ball of the induced RKHS space. Condition~\ref{new} requires this RKHS
unit ball to be stable under the pullback induced by each stochastic reverse
transition. Specifically, for an output-space RKHS unit-ball function \(f\),
\(\ell_{s,f}^x(\boldsymbol{y})=\mathbb E_{\varepsilon_s}
\{f(g_{(s)}^*(\boldsymbol{x},\boldsymbol{y},\boldsymbol{\varepsilon}_s))\}\) is the pulled-back function on the input
space. The condition states that, after centering, this function has a uniformly
bounded input-space RKHS norm. Intuitively, the transition should not turn
Energy-Distance RKHS functions into overly irregular functions of the previous
state. 
This condition holds for a broad class of transitions, including common Gaussian and certain nonlinear transformations; see
the Supplementary Material for further discussion and examples.
%for scalar affine Gaussian transitions, including
%linear Gaussian diffusion-type updates as special cases, and for certain one-dimensional monotone nonlinear and small residual Gaussian transitions; see the Supplementary Material for verification.
\end{remark}

In the Reverse Markov Engression setting, let
$\mathcal S=\cup_{s=1}^S \mathcal S_s$
denote the overall training randomness across \(S\) reverse steps, where
$
\mathcal S_s=
\{(\boldsymbol X_i,\boldsymbol Y_i^{s-1}, \boldsymbol Y_i^{s})\}_{i=1}^n
\cup
\{\boldsymbol\varepsilon_{ij}^{s}\}_{i=1,\dots,n;\,j=1,\dots,m}
$
is the training sample for the \(s\)th step generator \(\hat g_{(s)}\). Recall that \(P_{g^*}(\cdot\mid \boldsymbol X)\) denotes the target conditional distribution. 
In this subsection, \(P_{\hat g}(\cdot\mid \boldsymbol X)\) denotes the final conditional distribution induced by the fitted reverse Markov Engression procedure. Together with the error propagation result in Lemma 2 of the Supplementary Material, we obtain the following global convergence rate.

\begin{theorem}\label{thm:global_convergence}
    Suppose Conditions \ref{cond:boundedness_markov}-\ref{new} hold for each $s = 1, \dots, S$. For the generator $g_{(s)}^*$, let $d_{in} = p+q+d$ be the concatenated input dimension. Define $ \alpha_s = \beta_s/\{2(d_{in} + 2\beta_s)\}.$
    We have
    \begin{equation*}
\mathbb E_{\mathcal S}\mathbb E_{\boldsymbol X}
\left[
\mathcal D\big(
P_{\hat{g}}(\cdot\mid \boldsymbol X),
P_{g^*}(\cdot\mid \boldsymbol X)
\big)
\right]
\le
\mathcal C_S
n^{-\min_s\alpha_s}
\log(n)^{\max_s\left(\frac{\beta_s}{d_{in}}+\frac{\beta_s\vee 1}{2}\right)},
    \end{equation*}
    where $\mathcal{C}_S$ is a constant independent of $n$.% and $M_s = \prod_{k=1}^{s-1} L_k$ is the cumulative Lipschitz constant from step $s-1$ to $0$.
\end{theorem}
\begin{remark}
Theorem \ref{thm:global_convergence} shows that the reverse Markov Engression framework preserves the same polynomial convergence order as the single-step Engression result, with the overall rate determined by the least smooth reverse transition. Moreover, the cumulative Energy-Distance stability constants are absorbed into \(\mathcal C_S\); hence the reverse chain does not introduce rate deterioration beyond the hardest local step.
\end{remark}

% \begin{remark}
% Under this normalization, the multi-step rate achieves the same rate form
% as in the single-step setting, with the convergence exponent determined by
% the minimum smoothness across steps.
% This result aligns with the intuition that sequential generation does not
% deteriorate the statistical rate beyond the least smooth component,
% provided the transitions are Lipschitz stable. The theorem therefore
% provides theoretical support for the reverse Markov framework as a
% principled approach to multi-step conditional distribution learning.
% Notably, the cumulative Lipschitz constants
% $\prod_{k=1}^{s-1} L_k$ do not explicitly appear in the final rate, as they
% are absorbed into the constant $\mathcal{C}_S$ under Condition
% \ref{cond:reverse_lipschitz}.
% \end{remark}
% \begin{remark}
%     Comparing with Theorem \ref{rates} for the single-step setting, the multi-step rate achieves the same rate form but with the exponent determined by the minimum smoothness across steps. This result aligns with the intuition that sequential generation does not amplify the rate beyond the worst-case component, provided the transitions are Lipschitz stable. The theorem thus provides theoretical validation for the reverse Markov framework as a principled approach to multi-step conditional distribution learning.
% Notably, the cumulative Lipschitz constants \( \prod_{k=1}^{s-1} L_k\) do not explicitly appear in the final rate, as they are absorbed into the constant \(\mathcal{C}_S\) under Condition \ref{cond:reverse_lipschitz}.

% \end{remark}
\section{Conclusion and discussion}
In this paper, we provide theoretical guarantees for Engression and its
Reverse Markov extension under the Energy Distance. Our analysis connects the
energy-based distributional objective with finite-sample guarantees for deep
ReLU generators, and establishes nonasymptotic convergence rates for both
single-step Engression and multi-step reverse Markov learning.

Several directions remain open. First, although our rates are near-optimal in
terms of the ambient dimension, they still suffer from the curse of
dimensionality. Incorporating sufficient dimension reduction or other intrinsic
low-dimensional structure may lead to bounds depending on the effective
dimension rather than the nominal input dimension. Second, our analysis focuses
on ReLU networks; extending the theory to other architectures is an important
direction. Third, it would be valuable to generalize the framework beyond
Euclidean vector data, including functional, manifold-valued, or graph-structured
responses, where the Energy Distance and the generator class must be formulated
on appropriate mathematical spaces. Finally, our results are stated for
conditional distribution learning on the training support. Combining the present
finite-sample Energy-Distance guarantees with structural extrapolability
conditions, such as preadditive-noise models, may lead to out-of-support
distributional guarantees for Engression-type estimators.

\bibliographystyle{plainnat}
\bibliography{reference}

@article{tian2024reliable,
  title={Reliable generation of privacy-preserving synthetic electronic health record time series via diffusion models},
  author={Tian, Muhang and Chen, Bernie and Guo, Allan and Jiang, Shiyi and Zhang, Anru R},
  journal={Journal of the American Medical Informatics Association},
  volume={31},
  number={11},
  pages={2529--2539},
  year={2024},
  publisher={Oxford University Press}
}

@article{shen2025reverse,
  title = {{Reverse Markov Learning}: Multi-Step Generative Models for Complex Distributions},
  author={Shen, Xinwei and Meinshausen, Nicolai and Zhang, Tong},
  journal={arXiv preprint arXiv:2502.13747},
  year={2025}
}

@inproceedings{sohl2015deep,
  title={Deep unsupervised learning using nonequilibrium thermodynamics},
  author={Sohl-Dickstein, Jascha and Weiss, Eric and Maheswaranathan, Niru and Ganguli, Surya},
  booktitle = {International Conference on Machine Learning},
  pages={2256--2265},
  year={2015},
  organization={PMLR}
}

@article{ho2020denoising,
  title={Denoising diffusion probabilistic models},
  author={Ho, Jonathan and Jain, Ajay and Abbeel, Pieter},
  journal = {Advances in Neural Information Processing Systems},
  volume={33},
  pages={6840--6851},
  year={2020}
}

@article{shen2025engression,
  title={Engression: extrapolation through the lens of distributional regression},
  author={Shen, Xinwei and Meinshausen, Nicolai},
  journal={Journal of the Royal Statistical Society Series B: Statistical Methodology},
  volume={87},
  number={3},
  pages={653--677},
  year={2025},
  publisher={Oxford University Press UK}
}

@article{zhou2023deep,
  title={A deep generative approach to conditional sampling},
  author={Zhou, Xingyu and Jiao, Yuling and Liu, Jin and Huang, Jian},
  journal={Journal of the American Statistical Association},
  volume={118},
  number={543},
  pages={1837--1848},
  year={2023},
  publisher={Taylor \& Francis}
}

@article{yang2025frugal,
  title={Frugal, Flexible, Faithful: Causal Data Simulation via Frengression},
  author={Yang, Linying and Evans, Robin J and Shen, Xinwei},
  journal={arXiv preprint arXiv:2508.01018},
  year={2025}
}

@article{kraft2026modeling,
  title={Modeling uncertainty with engression: A deep generative time-series approach},
  author={Kraft, Basil and Stalder, Steven and Aeberhard, William H and Ruiz, Nicol{\'a}s Harrington and Meinshausen, Nicolai and Shen, Xinwei and Gudmundsson, Lukas},
  journal={Geophysical Research Letters},
  volume={53},
  number={2},
  pages={e2025GL120122},
  year={2026},
  publisher={Wiley Online Library}
}

@article{de2025distributional,
  title={Distributional diffusion models with scoring rules},
  author={De Bortoli, Valentin and Galashov, Alexandre and Guntupalli, J Swaroop and Zhou, Guangyao and Murphy, Kevin and Gretton, Arthur and Doucet, Arnaud},
  journal={arXiv preprint arXiv:2502.02483},
  year={2025}
}

@article{song2026wasserstein,
  title={Wasserstein generative regression},
  author={Song, Shanshan and Wang, Tong and Shen, Guohao and Lin, Yuanyuan and Huang, Jian},
  journal={Journal of the Royal Statistical Society Series B: Statistical Methodology},
  volume={88},
  number={1},
  pages={330--351},
  year={2026},
  publisher={Oxford University Press UK}
}

@article{song2019generative,
  title={Generative modeling by estimating gradients of the data distribution},
  author={Song, Yang and Ermon, Stefano},
 journal = {Advances in Neural Information Processing Systems},
  volume={32},
  year={2019}
}

@inproceedings{pinaya2022brain,
  title={Brain imaging generation with latent diffusion models},
  author={Pinaya, Walter HL and Tudosiu, Petru-Daniel and Dafflon, Jessica and Da Costa, Pedro F and Fernandez, Virginia and Nachev, Parashkev and Ourselin, Sebastien and Cardoso, M Jorge},
  booktitle={MICCAI workshop on deep generative models},
  pages={117--126},
  year={2022},
  organization={Springer}
}

@article{gao2020learning,
  title={Learning energy-based models by diffusion recovery likelihood},
  author={Gao, Ruiqi and Song, Yang and Poole, Ben and Wu, Ying Nian and Kingma, Diederik P},
  journal={arXiv preprint arXiv:2012.08125},
  year={2020}
}

@article{chen2020distribution,
  title={Distribution approximation and statistical estimation guarantees of generative adversarial networks},
  author={Chen, Minshuo and Liao, Wenjing and Zha, Hongyuan and Zhao, Tuo},
  journal={arXiv preprint arXiv:2002.03938},
  year={2020}
}

@inproceedings{zhou2025error,
  title={An error analysis of flow matching for deep generative modeling},
  author={Zhou, Zhengyu and Liu, Weiwei},
  booktitle={Forty-second International Conference on Machine Learning},
  year={2025}
}

@article{schmidt2020nonparametric,
  title = {Nonparametric Regression Using Deep Neural Networks with {ReLU} Activation Function},
  author={Schmidt-Hieber, Johannes},
  journal={The Annals of Statistics},
  volume={48},
  number={4},
  pages={1875},
  year={2020},
  publisher={Institute of Mathematical Statistics}
}

@article{han2022card,
  title={Card: Classification and regression diffusion models},
  author={Han, Xizewen and Zheng, Huangjie and Zhou, Mingyuan},
  journal={Advances in Neural Information Processing Systems},
  volume={35},
  pages={18100--18115},
  year={2022}
}

@article{su2026wasserstein,
  title={Wasserstein {GAN}-based estimation for conditional distribution function with current status data},
  author={Su, Wen and Liu, Changyu and Yin, Guosheng and Huang, Jian},
  journal={Lifetime Data Analysis},
  volume={32},
  number={1},
  pages={12},
  year={2026},
  publisher={Springer}
}

@article{gneiting2007strictly,
  title={Strictly proper scoring rules, prediction, and estimation},
  author={Gneiting, Tilmann and Raftery, Adrian E},
 journal = {Journal of the American Statistical Association},
  volume={102},
  number={477},
  pages={359--378},
  year={2007},
  publisher={Taylor \& Francis}
}

@book{gyorfi2002distribution,
  title     = {A Distribution-Free Theory of Nonparametric Regression},
  author    = {Gy{\"o}rfi, L{\'a}szl{\'o} and Kohler, Michael and Krzy{\.z}ak, Adam and Walk, Harro},
  year      = {2002},
  publisher = {Springer},
  series    = {Springer Series in Statistics}
}

@article{sejdinovic2013equivalence,
  title   = {Equivalence of Distance-Based and {RKHS}-Based Statistics in Hypothesis Testing},
  author  = {Sejdinovic, Dino and Sriperumbudur, Bharath and Gretton, Arthur and Fukumizu, Kenji},
  journal = {The Annals of Statistics},
  volume  = {41},
  number  = {5},
  pages   = {2263--2291},
  year    = {2013}
}

@article{szekely2013energy,
  title={Energy statistics: A class of statistics based on distances},
  author={Sz{\'e}kely, G{\'a}bor J and Rizzo, Maria L},
  journal={Journal of Statistical Planning and Inference},
  volume={143},
  number={8},
  pages={1249--1272},
  year={2013},
  publisher={Elsevier}
}

@inproceedings{bottou2018geometrical,
  title={Geometrical insights for implicit generative modeling},
  author={Bottou, Leon and Arjovsky, Martin and Lopez-Paz, David and Oquab, Maxime},
  booktitle={Braverman Readings in Machine Learning. Key Ideas from Inception to Current State: International Conference Commemorating the 40th Anniversary of Emmanuil Braverman's Decease, Boston, MA, USA, April 28-30, 2017, Invited Talks},
  pages={229--268},
  year={2018},
  organization={Springer}
}

\end{document}